\documentclass[twocolumn]{aastex61}
\usepackage{graphicx}				
\usepackage{amssymb}
\usepackage{color}
\usepackage{amsmath}
\usepackage[bottom]{footmisc}

\usepackage{natbib}
\newcommand{\zfourge}{{\sc zfourge}}
\newcommand{\Hbeta}{H$\beta$}
\newcommand{\OIII}{[\hbox{{\rm O}\kern 0.1em{\sc iii}}]}
\newcommand{\Ks}{$K_s$}
\newcommand{\Av}{A$_{\rm V}$}

\shortauthors{Forrest et al.}

\begin{document}

\title{Discovery of Extreme \OIII+\Hbeta\ Emitting Galaxies Tracing an Overdensity at $\MakeLowercase{z}\sim3.5$ in CDF-South\footnote{This paper includes data gathered with the 6.5 meter Magellan Telescopes located at Las Campanas Observatory, Chile.}}

\correspondingauthor{Ben Forrest}
\email{bforrest@physics.tamu.edu}

\author[0000-0001-6003-0541]{Ben Forrest}
\affiliation{George P. and Cynthia W. Mitchell Institute for Fundamental Physics and Astronomy, Department of Physics and Astronomy, Texas A\&M University, College Station, TX 77843, USA}

\author{Kim-Vy H. Tran}
\affiliation{George P. and Cynthia W. Mitchell Institute for Fundamental Physics and Astronomy, Department of Physics and Astronomy, Texas A\&M University, College Station, TX 77843, USA}

\author{Adam Broussard}
\affiliation{Department of Physics and Astronomy, Rutgers, The State University of New Jersey, 136 Frelinghuysen Road, Piscataway, NJ 08854, USA}

\author{Rebecca J. Allen}
\affiliation{Centre for Astrophysics and Supercomputing, Swinburne University, Hawthorn, VIC 3122, Australia}

\author{Miranda Apfel}
\affiliation{George P. and Cynthia W. Mitchell Institute for Fundamental Physics and Astronomy, Department of Physics and Astronomy, Texas A\&M University, College Station, TX 77843, USA}

\author{Michael J. Cowley}
\affiliation{Australian Astronomical Observatory, P.O. Box 915, North Ryde, NSW 1670, Australia}
\affiliation{Research Centre for Astronomy, Astrophysics \& Astrophotonics, Macquarie University, Sydney, NSW 2109, Australia}
\affiliation{Department of Physics \& Astronomy, Macquarie University, Sydney, NSW 2109, Australia}

\author{Karl Glazebrook}
\affiliation{Centre for Astrophysics and Supercomputing, Swinburne University, Hawthorn, VIC 3122, Australia}

\author{Glenn G. Kacprzak}
\affiliation{Centre for Astrophysics and Supercomputing, Swinburne University, Hawthorn, VIC 3122, Australia}

\author{Ivo Labb\'e}
\affiliation{Leiden Observatory, Leiden University, P.O. Box 9513, 2300 RA Leiden, The Netherlands}

\author{Themiya Nanayakkara}
\affiliation{Centre for Astrophysics and Supercomputing, Swinburne University, Hawthorn, VIC 3122, Australia}

\author{Casey Papovich}
\affiliation{George P. and Cynthia W. Mitchell Institute for Fundamental Physics and Astronomy, Department of Physics and Astronomy, Texas A\&M University, College Station, TX 77843, USA}

\author{Ryan F. Quadri}
\affiliation{George P. and Cynthia W. Mitchell Institute for Fundamental Physics and Astronomy, Department of Physics and Astronomy, Texas A\&M University, College Station, TX 77843, USA}

\author{Lee R. Spitler}
\affiliation{Australian Astronomical Observatory, P.O. Box 915, North Ryde, NSW 1670, Australia}
\affiliation{Research Centre for Astronomy, Astrophysics \& Astrophotonics, Macquarie University, Sydney, NSW 2109, Australia}
\affiliation{Department of Physics \& Astronomy, Macquarie University, Sydney, NSW 2109, Australia}

\author{Caroline M. S. Straatman}
\affiliation{Max Planck Institute for Astronomy, K\"onigstuhl 17, 69117 Heidelberg, Germany}

\author{Adam Tomczak}
\affiliation{Department of Physics, UC Davis, Davis, CA 95616, USA}

\begin{abstract}
Using deep multi-wavelength photometry of galaxies from \zfourge, we group galaxies at $2.5<z<4.0$ by the shape of their spectral energy distributions (SEDs).
We identify a population of galaxies with excess emission in the \Ks-band, which corresponds to \OIII+\Hbeta\ emission at $2.95<z<3.65$.
This population includes 78\% of the bluest galaxies with UV slopes steeper than $\beta = -2$.
We de-redshift and scale this photometry to build two composite SEDs, enabling us to measure equivalent widths of these Extreme \OIII+\Hbeta\ Emission Line Galaxies (EELGs) at $z\sim3.5$.
We identify 60 galaxies that comprise a composite SED with \OIII+\Hbeta\ rest-frame equivalent width of $803\pm228$\AA\ and another 218 galaxies in a composite SED with equivalent width of $230\pm90$\AA.
These EELGs are analogous to the `green peas' found in the SDSS, and are thought to be undergoing their first burst of star formation due to their blue colors ($\beta < -1.6$), young ages ($\log(\rm{age}/yr)\sim7.2$), and low dust attenuation values.
Their strong nebular emission lines and compact sizes (typically $\sim1.4$ kpc) are consistent with the properties of the star-forming galaxies possibly responsible for reionizing the universe at $z>6$.
Many of the EELGs also exhibit Lyman-$\alpha$ emission.
Additionally, we find that many of these sources are clustered in an overdensity in the Chandra Deep Field South, with five spectroscopically confirmed members at $z=3.474 \pm 0.004$.
The spatial distribution and photometric redshifts of the \zfourge\ population further confirm the overdensity highlighted by the EELGs.
\end{abstract}


\section{Introduction \& Background}

The discovery of galaxies with strong \OIII$\lambda$5007 emission, extreme star formation rates (SFRs), low masses, and low reddening by \cite{Cardamone2009} was key to finding objects thought to be responsible for the reionization of the Universe \citep{Robertson2013, Nakajima2014, Robertson2015}.
Known as `green peas' for their strong emission in the rest-frame optical and compact sizes, similar galaxies have since been found at much higher redshifts.
These green peas seem in many aspects to be more extreme versions of blue compact dwarfs \citep{Sargent1970}, having low masses and strong nebular emission lines, albeit with distinctly higher specific star formation rates \citep[sSFRs; e.g.,][]{Maseda2014}.

These objects exhibit bright optical nebular emission lines such as H$\alpha$ and \OIII, with rest-frame equivalent widths of several hundred to over one thousand angstroms, indicating bursty star forming activity \citep[e.g.,][hereafter S16]{Atek2011, VanderWel2011, Maseda2013, Stark2014, Sanders2016}.
Such galaxies are increasingly common at higher redshifts, appear to have enhanced \OIII$\lambda$5007 relative to both \Hbeta\ and [OII]$\lambda$3727, and are often Lyman-$\alpha$ Emitters (LAEs) \citep[e.g.,][]{Labbe2013, Smit2014, Holden2016, Trainor2016, Nakajima2016}.

While the strong emission lines of these objects can be detected with spectroscopy at $z\sim3$ \citep[][S16]{Holden2016, Nakajima2016}, the rest-frame optical stellar continuum is quite faint ($K_s>25$).
Indeed for such low-mass objects, this continuum has until recently been too faint for spectroscopy of large samples at these redshifts \citep[e.g.,][]{Nanayakkara2016}.
However, we can use the deep multi-wavelength photometry available in \zfourge\ to construct composite SEDs \citep{Forrest2016} and analyze the faint stellar continuum of a sample of emission-line objects at $2.5<z<4.0$.

We will refer to the most intense of these compact, \OIII-emitting galaxies, as Extreme Emission Line Galaxies (EELGs), and galaxies with less intense, but still significant \OIII\ emission as Strong Emission Line Galaxies (SELGs) for the duration of the Letter.
We assume a $\Lambda$CDM cosmology of $\Omega_M = 0.3$, $\Omega_\Lambda = 0.7$,
and $H_0 = 70$ km s$^{-1}$ Mpc$^{-1}$. 
For such a Universe, 1" = 7.320 kpc at $z=3.5$.
We use a Chabrier Initial Mass Function \citep{Chabrier2003} and the AB magnitude system \citep{Oke1983}.

\section{Data \& Methods} \label{Methods}

\subsection{[OIII] Emitter Photometry} \label{phot}

We use data from the medium band near-infrared FourStar Galaxy Evolution Survey \citep[\zfourge;][]{Straatman2016}.
This survey combines imaging from a large number of previous surveys to create the deepest \Ks-band detection image (see \citet{Straatman2016} for details) of the Chandra Deep Field South \citep[CDFS][]{Giacconi2002}), with a $5\sigma$ limiting depth of 26.5 AB mag. 
Imaging in the COSMOS \citep{Scoville2007} and UDS \citep{Lawrence2007} fields are also quite deep, at 25.5 and 25.7 magnitudes, respectively.
In addition, due to its medium band filters bracketing well-known features, such as the Balmer break, \OIII\ emission, and H$\alpha$ emission, \zfourge\ provides precise photometric redshifts.
Extensive testing against spectroscopic redshifts yielded $\sigma_{z,spec} = 0.01$ \citep{Nanayakkara2016,Straatman2016}.
The photometric redshift errors for objects without spectroscopic redshifts were also estimated from the photometric redshift differences of close pairs \citep{Quadri2010}.
For blue star-forming galaxies, this method yields $\sigma_{z,pairs} = 0.015$ at $z\sim2.5$, which agrees with the photo-$z$ determinations from EAZY \citep{Brammer2008}.

Forrest, et al., in prep, build a set of composite SEDs from non-AGN \zfourge\ galaxies in $2.5<z<4.0$, with $SNR_{K_s}>20$.
AGN are removed using the catalogs from \cite{Cowley2016}, which account for infrared, X-ray, and radio selected sources.
The composite SED method uses de-redshifted photometry to group galaxies together by SED shape, then scales the photometry of similar galaxies, termed analogs, to trace the intrinsic SED with better spectral resolution than is possible with the photometry of a single galaxy.
Medians of these de-redshifted photometric points in wavelength bins are then taken to be the composite SED.
Each point has a custom composite SED filter constructed from a weighted combination of the underlying photometry, which is used in fitting programs such as FAST \citep{Kriek2009}.
This method was introduced by \citet{Kriek2011} and details are also provided in \citet{Forrest2016}.

The bluest composite SEDs show significant emission from \OIII+\Hbeta, corresponding to a sample of EELGs (60 galaxies), and the other to a population of SELGs (218 galaxies).
The emission lines fall into the $K_s$-band in the redshift range $2.95<z<3.65$, and this excess flux in the \Ks-band increases our confidence in the photometric redshifts, which are in strong agreement with the handful of spectroscopic redshifts available for this population.

Spectroscopic redshifts for a subset of galaxies in our sample are found by cross-matching with several works including ESO/GOODS spectroscopy from \textit{FORS2} \citep{Vanzella2009} and \textit{VIMOS}  \citep{Balestra2010}, \textit{3D-HST} \citep{Momcheva2016}, and the ZFIRE survey \citep{Nanayakkara2016}.
These samples total 37 SELGs and 7 EELGs with either grism or spectroscopic redshifts.
There are 17 emission line galaxies with \textit{HST/WFC3/G141} grism redshifts from \textit{3D-HST}, with median $|\Delta z|/z=0.04$, and 27 with rest-frame ultraviolet spectroscopic redshifts, with median $|\Delta z|/z=0.016$.


	\begin{figure}[tp]
	\centerline{\includegraphics[width=0.55\textwidth,trim=0in 0in 0in 0in, clip=true]{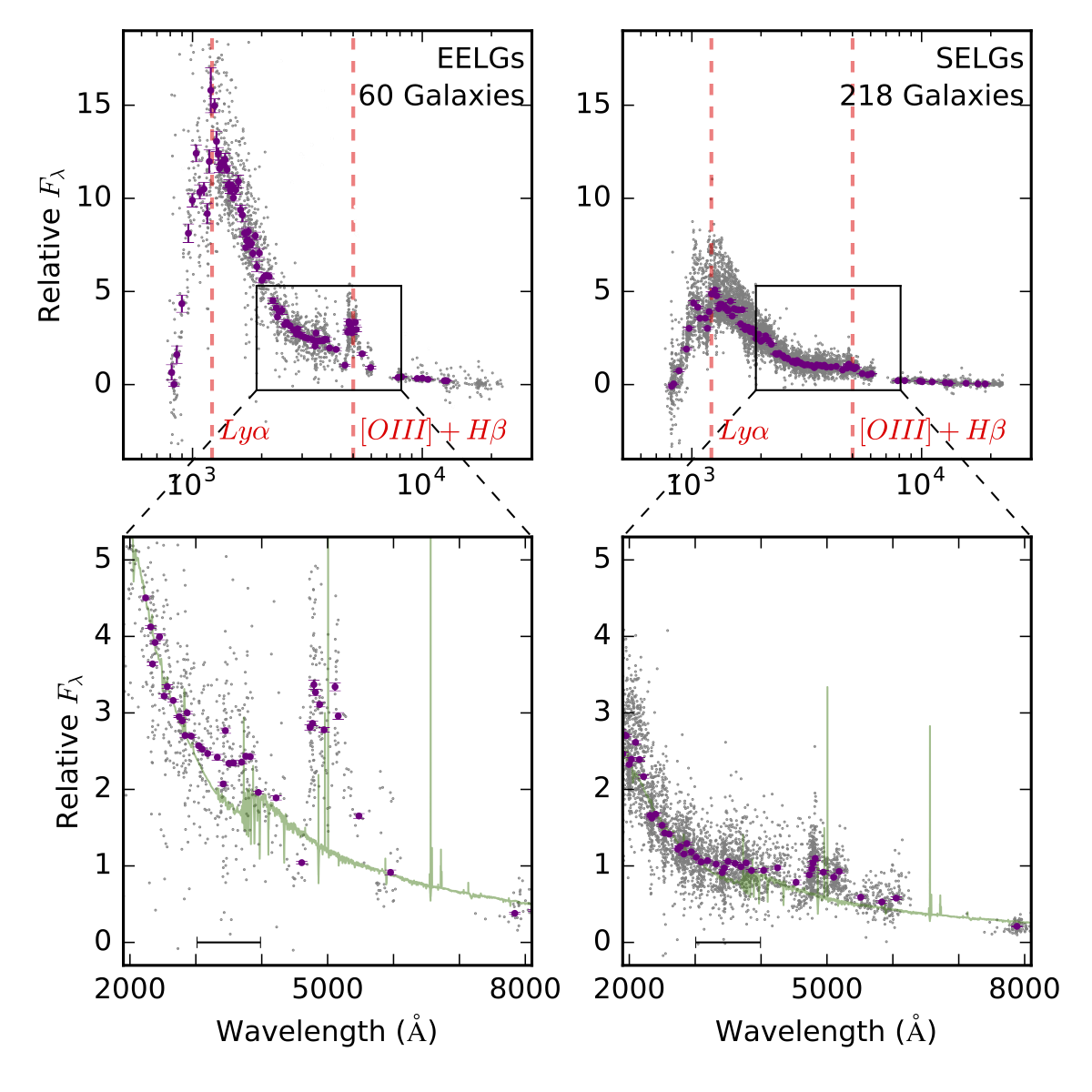}}
	\caption{ EELG (left) and SELG (right) composite SEDs.  These galaxies have significant emission from \OIII$\lambda$5007 + \Hbeta, in the \Ks-band at $z\sim3.5$. Their de-redshifted and scaled analog photometry is shown in gray, while the median points which make up the composite SEDs are in purple.  The best-fit model from FAST is green (emission lines are scaled down for clarity; see Section \ref{FitEF}), and the characteristic width of the composite filters is shown at the bottom. The strong emission in \OIII+\Hbeta\ is apparent and the EELG composite also shows Ly$\alpha$ emission.}
	\label{fig:comp}
	\end{figure}


\subsection{Fitting Stellar Continuum with FAST}  \label{FitEF}

We make use of FAST to fit the stellar continuum of the composite SEDs using the \cite{Bruzual2003} (BC03) models to obtain equivalent widths for optical emission lines in Section \ref{EW}.
Accounting for emission lines as strong as those observed in the EELGs is of critical importance during this process.
Failure to do so will result in FAST overestimating the stellar continuum, leading to errors in parameters such as stellar mass, SFR, and age \citep[e.g.][hereafter S15; Spitler, et al., in prep]{Erb2010, Atek2011, Reddy2012, Salmon2015}.

We test three methods for fitting both the composite SEDs and the individual galaxies using the BC03 models.
Declining-$\tau$ models for all three methods have $7<\log(\tau/\rm{yr})<11$, ages $6<\log(\rm{age/yr})<9.25$, and dust attenuations $0<$\Av/$\rm{mag}<4$.

First, we fit the BC03 models to the set of photometry (Method A).
The strength of the emission feature artificially raises the level of the stellar continuum, and thus the mass, for this method.
Secondly, we use the same models, but mask out points affected by \OIII+\Hbeta\ emission (Method B).
The third method uses a set of high resolution ($\Delta \lambda = 1$\AA) BC03 models with emission lines added fit to the entire composite SED (Method C).
These emission line models, as detailed in Section 3 of S15, include 119 sets of relative emission-line strengths with nebular emission taken into account. 
While the nebular continuum is not considered in this method, effects from this flux are minimal in the rest-frame optical, and so can be safely ignored (S15).
We also run sets of models with several metallicity values.

Method C yields the best fit to the composite SED, with $\chi^2=4.97$ for the EELG composite, compared to $\chi^2=7.83$ using Method A, which is used in the \zfourge\ catalogs.
The best fits also have a sub-solar metallicity, $Z=0.004$, in agreement with studies suggesting low metallicities for these galaxies \citep[e.g.][S15; S16]{Salzer2005, Izotov2011, VanderWel2011}.
The resulting mean values and one sigma spreads for the individual ELGs are shown in Table \ref{tab:table1}.

\section{[OIII]+\Hbeta\ Emission Line Galaxies}

\subsection{Galaxy Properties} \label{Prop}

Using Method C described above, we find that the EELG population in general is younger than the SELG population by 0.3 dex, has 0.5 dex less mass, and has more intense star formation (sSFR of 55 Gyr$^{-1}$ to 15 Gyr$^{-1}$).
This is consistent with a picture of the SELG population being on the same evolutionary path, but slightly more evolved.

We find mean physical sizes of 1.34 kpc and 1.63 kpc for the EELG and SELG populations by cross-matching with the $H_{F160W}$ sizes from CANDELS \citep[][$\lambda_{REST}\sim3400$\AA\ at $z\sim3.5$]{VanderWel2012}.
However, we note that at $z\sim3.5$ the angular resolution of $HST/WFC3/H_{F160W}\sim0.95$ kpc, and so a number of these galaxies are unresolved.
Nonetheless, the small sizes are consistent with measurements in both the local and distant Universe.
\cite{Henry2015} find a NUV Petrosian radius of $\sim1$ kpc at $z\sim0.2$ ($\lambda_{REST}\sim1900$\AA).
The $1.4<z<2.3$ sample from \cite{Maseda2014} has sizes ranging from $0.5<r_{eff}/kpc<1.6$, also based on data from \cite{VanderWel2012} ($4620<\lambda_{REST}/$\AA$<6350$).
These data suggest that EELG populations across cosmic time have the same physical sizes at rest-frame UV wavelengths.
Additionally, these sizes are consistent with the size-mass relation for $z\sim3.5$ derived in \citet{Allen2017}, which when extrapolated to $\log(M/M_\odot)\sim9$ predicts a size of 1.2 kpc.

One of the objects in our EELG sample is the subject of S16, which analyzes a rest-frame optical spectrum from MOSFIRE.
S16 finds galaxy properties in agreement with those derived in this work, namely a low mass ($\log(M_\star / M_\odot) = 9.33$), high sSFR (23 Gyr$^{-1}$), young age (160 Myr), and low dust attenuation ($E(B-V)_{stars}=0.12$).
Critically, they also find a metallicity of 12+(O/H) = 8.00, which confirms the low metallicities characteristic of galaxies undergoing early star formation \citep{Izotov2011}.

\begin{table*}[t]
  \centering
  \caption{Properties of the ELG Population}
  \label{tab:table1}
  \begin{tabular}{l | c c c | c c c}
      & & EELGs & & & SELGs & \\
      & Method A\footnote{All results are from FAST with a metallicity of $Z=0.004$.} & Method B & Method C\footnote{We reference values from Method C throughout the text.} & Method A & Method B & Method C\\
      \hline \hline
      $\log(M_*/M_\odot)$     & $9.2^{+0.3}_{-0.4}$ & $8.9^{+0.4}_{-0.3}$ & $8.7^{+0.2}_{-0.3}$    & $9.4^{+0.3}_{-0.3}$ & $9.1^{+0.3}_{-0.3}$ & $9.1^{+0.3}_{-0.3}$ \\
      $\log(\rm{age/yr})$     & $8.0^{+0.6}_{-1.2}$ & $7.6^{+0.6}_{-0.6}$ & $7.2^{+0.1}_{-0.3}$    & $8.2^{+0.2}_{-0.2}$ & $7.6^{+0.7}_{-0.6}$ & $7.5^{+0.4}_{-0.3}$ \\
      $\log(\tau \rm{/yr})$     & $8.1^{+0.6}_{-1.1}$ & $8.6^{+2.0}_{-1.6}$ & $7.7^{+1.0}_{-0.7}$    & $8.0^{+0.4}_{-0.6}$ & $8.2^{+1.0}_{-1.1}$ & $7.3^{+0.3}_{-0.3}$ \\
      SFR $(M_\odot \rm{/yr})$   & $21^{+25}_{-19}$ & $45^{+13}_{-40}$ & $24^{+15}_{-15}$    & $23^{+17}_{-21}$ & $50^{+32}_{-41}$ & $18^{+12}_{-12}$ \\
      \Av (mag)     & $0.21^{+0.44}_{-0.21}$ & $0.42^{+0.32}_{-0.31}$ & $0.45^{+0.20}_{-0.25}$    & $0.15^{+0.00}_{-0.15}$ & $0.53^{+0.27}_{-0.23}$ & $0.46^{+0.19}_{-0.21}$ \\
      \hline
      $EW_{[OIII] + H\beta}$\footnote{Rest-frame equivalent widths derived from composite SEDs.} (\AA) & $309\pm115$ & $659\pm189$ & $803\pm228$  & $100\pm67$ & $250\pm103$ & $230\pm90$ \\
      \hline
      $r_e$\footnote{From \citet{VanderWel2012} catalogs. At $z=3.5$, resolution is 0.95". This is independent of the FAST fitting.} (kpc) & & & $1.34^{+0.52}_{-0.74}$ & & & $1.62^{+0.65}_{-0.84}$ \\
   \end{tabular}
\end{table*}

\subsection{Large [OIII]+\Hbeta\ Equivalent Widths}\label{EW}

While the BC03+emission line models produce good fits to the composite SEDs (see Figure \ref{fig:comp}), they still underpredict the strength of the \OIII+\Hbeta\ emission feature.
To estimate the equivalent width, we remove the emission features from the model and add back in emission lines to match the observed composite SED feature.

We take the continuum to be the level of the best-fit spectrum from FAST using Method C where there are no emission lines, and use a simple linear interpolation to derive the continuum in areas with such features.
Since these models have resolution of 1\AA\ in the optical, all absorption features (i.e., \Hbeta\ absorption) are retained.
After obtaining the continuum fit to the composite SED, we add to it an emission line model with the following flux ratios: $F_{[OIII]\lambda5007} = 3 F_{[OIII]\lambda4959} = 7 F_{H\beta}$ \citep[e.g.][S16]{Salzer2005, VanderWel2011, Holden2016, Trainor2016}.
The resultant spectrum is convolved with the custom composite SED filters to create synthetic photometry.
The amplitude of the emission line model is varied to minimize the least squares of the composite SED and the synthetic photometry.
We do not have the resolution to separate the effects of the lines but ratios of \OIII$\lambda$5007 to \Hbeta\ from 1 to 20 were tested with negligible change to the resulting equivalent width.
Errors on the equivalent width were calculated by fitting a range of line strengths to form a grid of $\chi^2$ values and one sigma errors were calculated using the relation $P \propto e^{-\chi^2/2}$.

\begin{figure}[tp]
	\centerline{\includegraphics[width=0.5\textwidth,trim=0.25in 0in 0in 0in, clip=true]{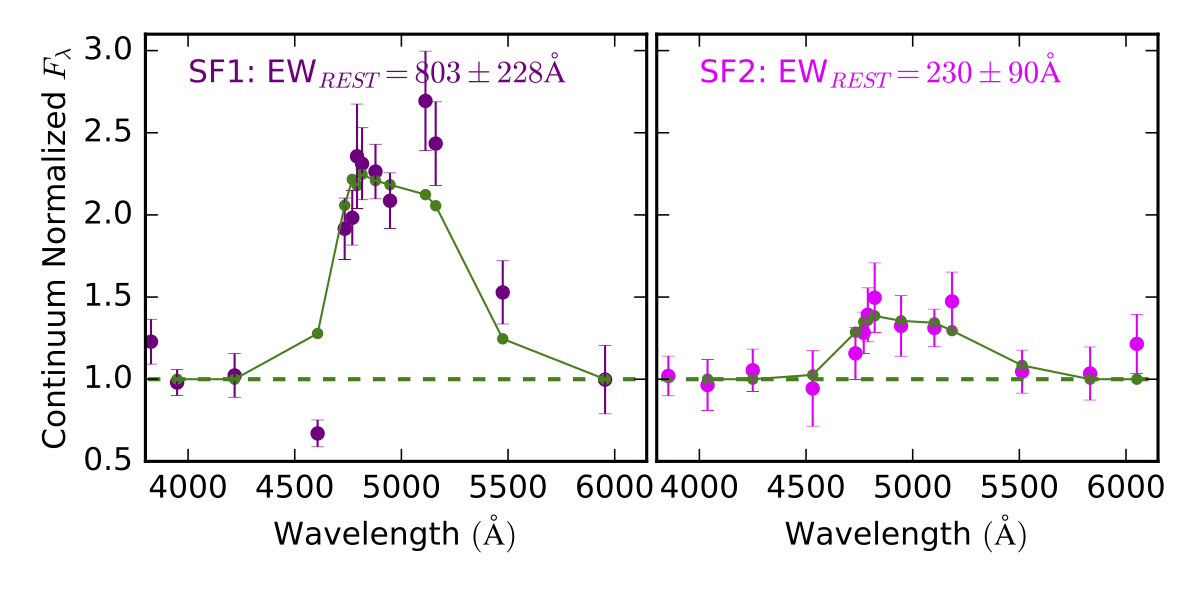}}
	\caption{The equivalent width for our two composite SEDs. The purple points show the composite SED normalized to the level of the stellar continuum as fit by FAST in the range while masking the emission feature.  Green points represent the synthetic photometry of the stellar continuum + emission line model in the effective composite SED filter curves.\\
	\\
	}
	\label{fig:ew}
	\end{figure}

For the EELG composite SED, we obtain a rest-frame equivalent width $EW_{[OIII]+ H\beta} = 803\pm228$\AA, corresponding to an observed-frame equivalent width $EW_{[OIII]+ H\beta} = 3565$\AA\ at the median redshift $z=3.44$ (see Figure \ref{fig:ew}).
The rest-frame equivalent width for the EELG population is consistent with those of EELGs in many studies which range from $\sim200-1600$\AA\ \citep[][S16]{Atek2011, VanderWel2011, Maseda2013}.
The SELG population has a rest-frame equivalent width $EW_{[OIII]+H\beta} = 230\pm90$\AA.
Spectroscopic follow-up of these objects would be useful for not only confirming the accuracy of this method, but also analyzing the spread of equivalent width values and \OIII/\Hbeta\ line ratios within the population.

To quantify uncertainties in the equivalent widths that are due to uncertainties in the photometric redshifts, we redshift a high equivalent width template and calculate synthetic $K_s$-band photometric values.  
We then mimic our composite SED method by de-redshifting the template using the `correct' redshift with noise added according to the $|\Delta z|/z$ values from spectroscopy (see Section \ref{phot}).  
Repeating this process 100 times, we measure the equivalent width to be $868\pm273$\AA, which matches very well with our measurement of $803\pm228$ \AA (Table 1, Method C).

We also note that both composite SEDs show emission in the ultraviolet (UV), consistent with many of the analogs having Ly$\alpha$ emission.
Several studies have suggested that LAEs are similar to the EELGs \citep[e.g.,][]{Hagen2016, Trainor2016, Nakajima2016}; our sample is consistent with that picture.


\subsection{Incidence of ELGs at High Redshift}

\OIII\ EELGs are less common locally than at higher redshifts \citep[e.g.,][]{Atek2011, VanderWel2011, Maseda2014, Maseda2016, Smit2014}.
In the redshift range of $2.5<z<4.0$ we detect 60 EELGs with $SNR_{K_s}>20$, with the faintest at $K_s=24.9$ magnitudes, and a characteristic equivalent width of $EW_{REST}=803$\AA.
Separating the EELGs by field we find $6.29 \times 10^{-5}$ Mpc$^{-3}$,  $0.897 \times 10^{-5}$ Mpc$^{-3}$,  and $2.15 \times 10^{-5}$ Mpc$^{-3}$ for CDFS, COSMOS, and UDS respectively.
SELGs have densities of $23.0 \times 10^{-5}$ Mpc$^{-3}$, $4.34 \times 10^{-5}$ Mpc$^{-3}$, and $7.02 \times 10^{-5}$ Mpc$^{-3}$ in the same order, and are 3.7 times more common than the EELGs in total.

In the redshift range $2.5<z<4.0$, these EELGs and SELGs dominate the population of the bluest galaxies with $\beta<-2$.
We select all ZFOURGE galaxies with $(U-V)_{REST}<1$, $(V-J)_{REST}<1$ and at least 5 photometric detections in the restframe wavelength range $1250<\lambda/\rm{\AA}<2600$, and fit a power law, $F_\lambda \propto \lambda^\beta$ to obtain the UV slope, $\beta$ \citep{Calzetti1994}.
Of the 58 such galaxies with $\beta<-2$, 45 (78\%) fall into our EELG or SELG groups, indicating that the majority of the bluest galaxies have strong emission features.

\section{Large Scale Structure in CDFS at $\MakeLowercase{z}\sim3.5$}

\begin{figure}[tp]
	\centerline{\includegraphics[width=0.5\textwidth,trim=0in 0in 0in 0in, clip=true]{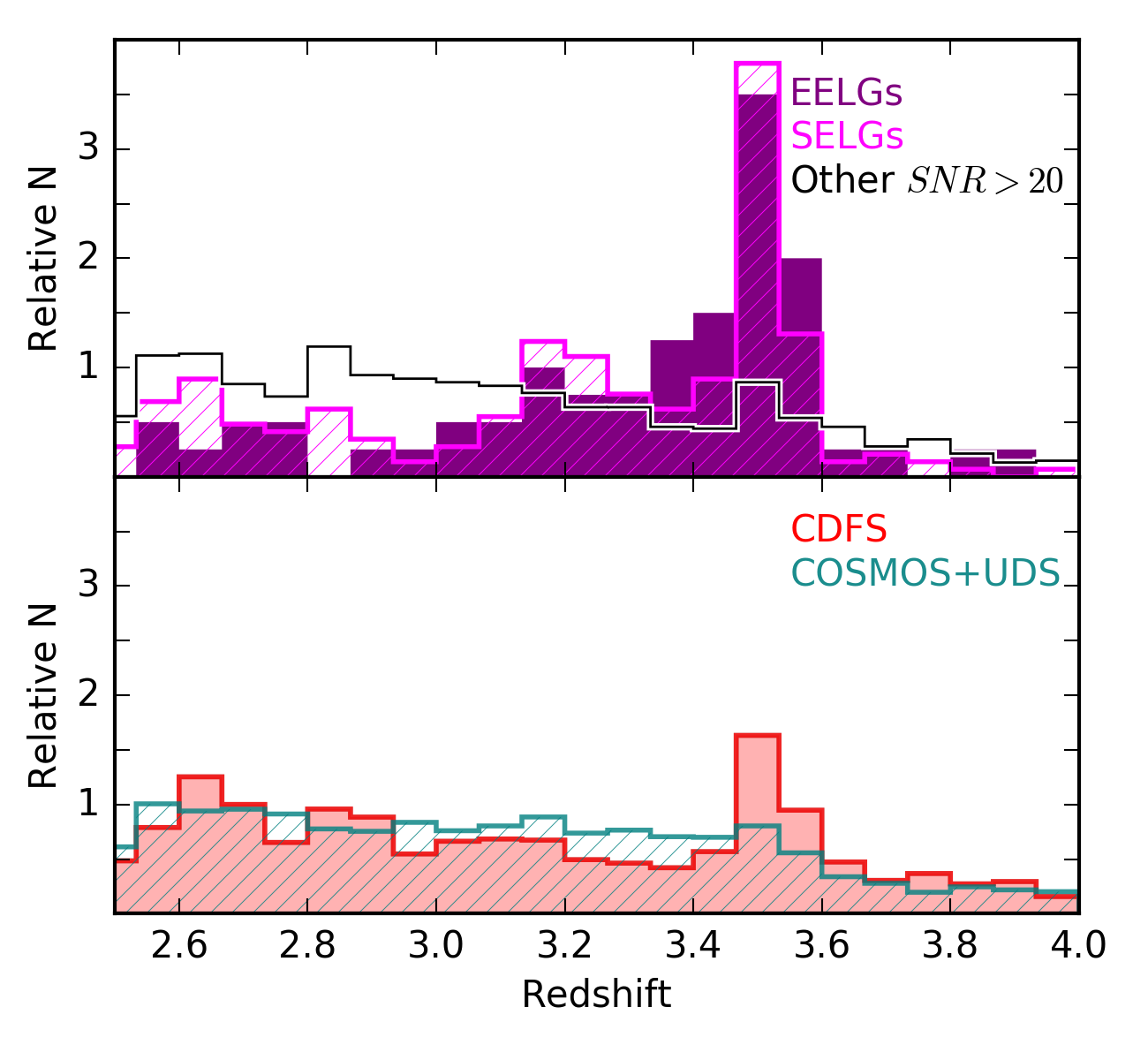}}
	\caption{ The top panel shows the photometric redshifts for the EELG and SELG populations in purple and light purple, respectively, as well as galaxies from the other composite SEDs of Forrest, et al., in prep, in black.  On the bottom is the normalized distribution of photometric redshifts for all galaxies with $K_s<24.9$ in $2.5<z<4$.
	}
	\label{fig:hist}
	\end{figure}

The EELGs and SELGs show an unexpected peak at $z\sim3.5$ in CDFS.
The full redshift distribution of galaxies in the ZFOURGE catalog, both photometric and spectroscopic, also confirms the presence of a galaxy overdensity at $z\sim3.5$ in CDFS (see Figure \ref{fig:hist}, as well as Figure 23 of \cite{Straatman2016}).
Comparing the CDFS redshift distribution to the combined distribution from COSMOS and UDS shows a strong difference, greater than $10\sigma$ as found using a K-S test.
Such an overdensity has been suggested by data from \textit{3D-HST} \citet[][Figure 24]{Skelton2014}, who use EAZY as well. 
The same is not found in CDF-North, which has similar filter coverage and $K_s$-band depth, and selects based on $HST$ $F125W+F140W+F160W$ imaging (their tables 6 and 7).
This removes the possibility of the overdensity being due to a bias in EAZY.


	\begin{figure*}[tp]
	\centerline{\includegraphics[width=\textwidth,trim=0in 0in 0in 0in, clip=true]{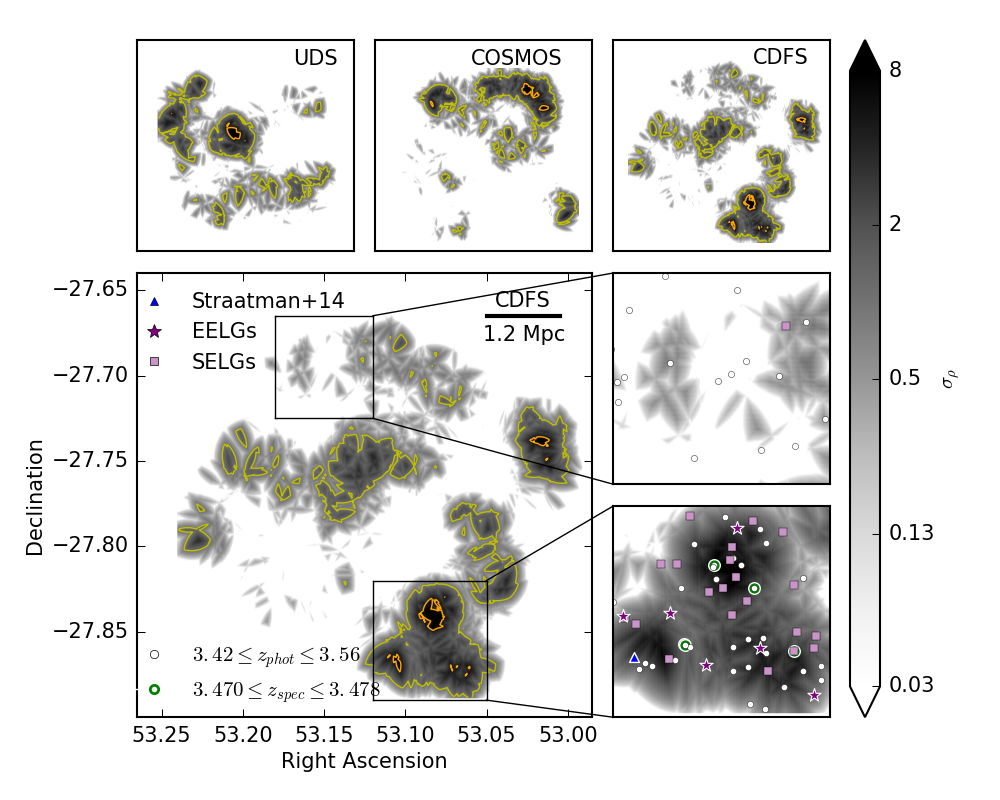}}
	\caption{\textit{Top:} Projected spatial density in all three \zfourge\ fields of the galaxies in the redshift range $3.42 \leq z \leq 3.56$ and brighter than $K_s<24.9$.  The density at a point is determined using the $7^{th}$ nearest neighbor metric, and shaded by number of standard deviations above the mean density in the field. Both $1\sigma$ (yellow) and $5\sigma$ (orange) contours are shown. \textit{Bottom:} Projected spatial density for CDFS with galaxies overplotted. Galaxies with $3.470 \leq z_{spec} \leq 3.478$ are shown in green and a massive quiescent galaxy from \cite{Straatman2014} at $z=3.56$ is displayed in blue.  For individual galaxies in  $3.42 \leq z \leq 3.56$, EELGs are purple stars, SELGs are magenta squares, other galaxies are white circles. Note that the overdense region has five spectroscopically confirmed galaxies - the leftmost two have a projected separation of $\sim 1$ arcsecond, and so are not resolved at this scale.}
	\label{fig:density}
	\end{figure*}


To map possible overdensities projected on the sky, we use the $7^{th}$ nearest-neighbor measure \citep[e.g.,][]{Papovich2010, Spitler2012}.
For each field, we create a grid of 1000 points on a side, and calculate the distance to the $7^{th}$ closest galaxy at each point, which is converted into a density.
Each point in all three fields is shaded by the number of standard deviations above the mean density in the respective field (see Figure \ref{fig:density}).
Additional density calculations were made using the $3^{rd}, 5^{th}$, and $10^{th}$ nearest neighbors - the results are similar to those from the $7^{th}$ nearest neighbor measure.

In the top row of Figure \ref{fig:density}, we show all three fields considering galaxies in the range $3.42<z<3.56$, with a magnitude cut of $K_s<24.9$ mag, where all three fields are complete \citep{Straatman2016}.
The 5 sigma peak in the UDS contains 10 galaxies while the 5 sigma peaks in COSMOS contain $10-30$ galaxies. 
In contrast, the largest overdensity in CDFS contains 53 galaxies. 
Slightly increasing or decreasing the redshift ranges does not change our results. 
We focus on the most significant overdensity at z~3.5 which is in CDFS but note that we do not exclude an overdensity in COSMOS.

At z=3.5, the galaxy overdensity in CDFS at (RA, Dec) = (53.08, -27.85) is approximately 1.8 Mpc in size.
Spectroscopy by \citet{Vanzella2009} in CDFS confirm five galaxies at z=3.474 (Figure \ref{fig:density}, green circles). 
\citet{Straatman2014} also report a massive quiescent galaxy at z=3.56 (Figure \ref{fig:density}, blue triangle).

Previous studies find that blue galaxies such as LAEs and LBGs trace large scale structure \citep[e.g.,][]{Steidel1998, Ouchi2005}.
Our results confirm that galaxies with strong \OIII + \Hbeta\ emission also exist in such overdensities.
EELGs have the added advantage that they can be spectroscopically confirmed more easily via their strong emission.

A possible bias in our result is that the nebular emission at $z\sim3.5$ is boosting the flux in the $K_s$-band such that we are preferentially detecting lower mass galaxies at this redshift.
However, the redshift and projected distributions of galaxies without these strong emission features show evidence of an overdensity as well.

\section{Conclusions}

We utilized multi-wavelength photometry from the \zfourge\ survey to build composite SEDs based on similar galaxy redshift and SED shape, revealing a population of galaxies with very blue colors and excess emission in the \Ks-band at $z\sim3.5$.
Parameters derived from FAST characterize these galaxies as having very young ages ($\log(\rm{age/yr})\sim7.2$), low masses ($\log(M/M_{\odot}) \sim 8.6$), and low dust content (\Av\ $\sim0.45$).
In addition, they are small in size ($r_e\sim 1.3$ kpc), and their remarkable emission from \OIII+\Hbeta, $EW_{REST}=803\pm228$\AA\ for the strongest emitters, is consistent with the properties of extreme star-forming galaxies that may have reionized the universe.

We observe that these EELGs appear on the order of $10^{-5}$ Mpc$^{-3}$ at $z\sim3.5$, and make up the vast majority (78\%) of the galaxies with $\beta<-2$ at this epoch.

Finally, we explored the distribution of these galaxies on the sky, and find an overdensity in the overall population of CDFS at $z\sim3.5$.
This giant structure is $\sim2$ Mpc in projected size and is a candidate progenitor of a galaxy cluster environment.
It also includes a massive quiescent galaxy from \cite{Straatman2014} and rest-frame UV spectroscopically confirmed LBGs \citep{Vanzella2009} in a very narrow redshift range.

Further spectroscopic follow-up of these galaxies is critical for understanding the earliest star-forming galaxies, and will also lead to interesting science cases for the next generation of telescopes, including JWST.
Such telescopes would be capable of detecting H$\alpha$ emission from these galaxies, which currently falls between the $K$-band and the \textit{Spitzer} 3.6$\mu m$ band, as well as looking for large scale structure based on optical emission lines.

\section*{Acknowledgments}

We wish to thank the Mitchell family, particularly the late George P. Mitchell, for their continuing support of astronomy.
We also thank the Carnegie Observatories and the Las Campanas Observatory for their
assistance in making the \zfourge\ survey possible.
B. Forrest and K. Tran acknowledge the support of the National Science Foundation under Grant \#1410728.
GGK acknowledges the support of the Australian Research Council through the award of a Future Fellowship (FT140100933).
Finally, we thank the anonymous referee for a series of comments which greatly improved the manuscript.


\begin{thebibliography}{}
\expandafter\ifx\csname natexlab\endcsname\relax\def\natexlab#1{#1}\fi

\bibitem[{Allen {et~al.}(2017)Allen, Kacprzak, Glazebrook, Labb{\'{e}}, Tran,
  Spitler, Cowley, Nanayakkara, Papovich, Quadri, Straatman, Tilvi, \& van
  Dokkum}]{Allen2017}
Allen, R.~J., Kacprzak, G.~G., Glazebrook, K., {et~al.} 2017, The Astrophysical
  Journal, 834, L11

\bibitem[{Atek {et~al.}(2011)Atek, Siana, Scarlata, Malkan, McCarthy, Teplitz,
  Henry, Colbert, Bridge, Bunker, Dressler, Fosbury, Hathi, Martin, Ross, \&
  Shim}]{Atek2011}
Atek, H., Siana, B., Scarlata, C., {et~al.} 2011, The Astrophysical Journal,
  743, 121

\bibitem[{Balestra {et~al.}(2010)Balestra, Mainieri, Popesso, Dickinson,
  Nonino, Rosati, Teimoorinia, Vanzella, Cristiani, Cesarsky, Fosbury,
  Kuntschner, \& Rettura}]{Balestra2010}
Balestra, I., Mainieri, V., Popesso, P., {et~al.} 2010, Astronomy and
  Astrophysics, 512, A12

\bibitem[{Brammer {et~al.}(2008)Brammer, van Dokkum, \& Coppi}]{Brammer2008}
Brammer, G.~B., van Dokkum, P.~G., \& Coppi, P. 2008, The Astrophysical
  Journal, 686, 1503

\bibitem[{Bruzual \& Charlot(2003)}]{Bruzual2003}
Bruzual, G., \& Charlot, S. 2003, Monthly Notices of the Royal Astronomical
  Society, 344, 1000

\bibitem[{Calzetti {et~al.}(1994)Calzetti, Kinney, \&
  Storchi-Bergmann}]{Calzetti1994}
Calzetti, D., Kinney, A.~L., \& Storchi-Bergmann, T. 1994, The Astrophysical
  Journal, 429, 582

\bibitem[{Cardamone {et~al.}(2009)Cardamone, Schawinski, Sarzi, Bamford,
  Bennert, Urry, Lintott, Keel, Parejko, Nichol, Thomas, Andreescu, Murray,
  Raddick, Slosar, Szalay, \& VandenBerg}]{Cardamone2009}
Cardamone, C., Schawinski, K., Sarzi, M., {et~al.} 2009, Monthly Notices of the
  Royal Astronomical Society, 399, 1191

\bibitem[{Chabrier(2003)}]{Chabrier2003}
Chabrier, G. 2003, Publications of the Astronomical Society of the Pacific,
  115, 763

\bibitem[{Cowley {et~al.}(2016)Cowley, Spitler, Tran, Rees, Labb{\'{e}}, Allen,
  Brammer, Glazebrook, Hopkins, Juneau, Kacprzak, Mullaney, Nanayakkara,
  Papovich, Quadri, Straatman, Tomczak, \& van Dokkum}]{Cowley2016}
Cowley, M.~J., Spitler, L.~R., Tran, K.-V.~H., {et~al.} 2016, Monthly Notices
  of the Royal Astronomical Society, 457, 629

\bibitem[{Erb {et~al.}(2010)Erb, Pettini, Shapley, Steidel, Law, \&
  Reddy}]{Erb2010}
Erb, D.~K., Pettini, M., Shapley, A.~E., {et~al.} 2010, The Astrophysical
  Journal, 719, 1168

\bibitem[{Forrest {et~al.}(2016)Forrest, Tran, Tomczak, Broussard, Labb{\'{e}},
  Papovich, Kriek, Allen, Cowley, Dickinson, Glazebrook, van Houdt, Inami,
  Kacprzak, Kawinwanichakij, Kelson, McCarthy, Monson, Morrison, Nanayakkara,
  Persson, Quadri, Spitler, Straatman, \& Tilvi}]{Forrest2016}
Forrest, B., Tran, K.-V.~H., Tomczak, A.~R., {et~al.} 2016, The Astrophysical
  Journal, 818, L26

\bibitem[{Giacconi {et~al.}(2002)Giacconi, Zirm, Wang, Rosati, Nonino, Tozzi,
  Gilli, Mainieri, Hasinger, Kewley, Bergeron, Borgani, Gilmozzi, Grogin,
  Koekemoer, Schreier, Zheng, \& Norman}]{Giacconi2002}
Giacconi, R., Zirm, A., Wang, J., {et~al.} 2002, The Astrophysical Journal
  Supplement Series, 139, 369

\bibitem[{Hagen {et~al.}(2016)Hagen, Zeimann, Behrens, Ciardullo, Gebhardt,
  Gronwall, Bridge, Fox, Schneider, Trump, Blanc, Chiang, Chonis, Finkelstein,
  Hill, Jogee, \& Gawiser}]{Hagen2016}
Hagen, A., Zeimann, G.~R., Behrens, C., {et~al.} 2016, The Astrophysical
  Journal, 817, 79

\bibitem[{Henry {et~al.}(2015)Henry, Scarlata, Martin, \& Erb}]{Henry2015}
Henry, A., Scarlata, C., Martin, C.~L., \& Erb, D. 2015, The Astrophysical
  Journal, 809, 19

\bibitem[{Holden {et~al.}(2016)Holden, Oesch, Gonz{\'{a}}lez, Illingworth,
  Labb{\'{e}}, Bouwens, Franx, van Dokkum, \& Spitler}]{Holden2016}
Holden, B.~P., Oesch, P.~A., Gonz{\'{a}}lez, V.~G., {et~al.} 2016, The
  Astrophysical Journal, 820, 73

\bibitem[{Izotov {et~al.}(2011)Izotov, Guseva, \& Thuan}]{Izotov2011}
Izotov, Y.~I., Guseva, N.~G., \& Thuan, T.~X. 2011, The Astrophysical Journal,
  728, 161

\bibitem[{Kriek {et~al.}(2009)Kriek, van Dokkum, Labb{\'{e}}, Franx,
  Illingworth, Marchesini, \& Quadri}]{Kriek2009}
Kriek, M., van Dokkum, P.~G., Labb{\'{e}}, I., {et~al.} 2009, The Astrophysical
  Journal, 700, 221

\bibitem[{Kriek {et~al.}(2011)Kriek, van Dokkum, Whitaker, Labb{\'{e}}, Franx,
  \& Brammer}]{Kriek2011}
Kriek, M., van Dokkum, P.~G., Whitaker, K.~E., {et~al.} 2011, The Astrophysical
  Journal, 743, 168

\bibitem[{Labb{\'{e}} {et~al.}(2013)Labb{\'{e}}, Oesch, Bouwens, Illingworth,
  Magee, Gonz{\'{a}}lez, Carollo, Franx, Trenti, van Dokkum, \&
  Stiavelli}]{Labbe2013}
Labb{\'{e}}, I., Oesch, P.~a., Bouwens, R.~J., {et~al.} 2013, The Astrophysical
  Journal, 777, L19

\bibitem[{Lawrence {et~al.}(2007)Lawrence, Warren, Almaini, Edge, Hambly,
  Jameson, Lucas, Casali, Adamson, Dye, Emerson, Foucaud, Hewett, Hirst,
  Hodgkin, Irwin, Lodieu, McMahon, Simpson, Smail, Mortlock, \&
  Folger}]{Lawrence2007}
Lawrence, a., Warren, S.~J., Almaini, O., {et~al.} 2007, Monthly Notices of the
  Royal Astronomical Society, 379, 1599

\bibitem[{Maseda {et~al.}(2013)Maseda, van~der Wel, da~Cunha, Rix, Pacifici,
  Momcheva, Brammer, Franx, van Dokkum, Bell, Fumagalli, Grogin, Kocevski,
  Koekemoer, Lundgren, Marchesini, Nelson, Patel, Skelton, Straughn, Trump,
  Weiner, Whitaker, \& Wuyts}]{Maseda2013}
Maseda, M.~V., van~der Wel, A., da~Cunha, E., {et~al.} 2013, The Astrophysical
  Journal, 778, L22

\bibitem[{Maseda {et~al.}(2014)Maseda, van~der Wel, Rix, da~Cunha, Pacifici,
  Momcheva, Brammer, Meidt, Franx, van Dokkum, Fumagalli, Bell, Ferguson,
  F{\"{o}}rster-Schreiber, Koekemoer, Koo, Lundgren, Marchesini, Nelson, Patel,
  Skelton, Straughn, Trump, \& Whitaker}]{Maseda2014}
Maseda, M.~V., van~der Wel, A., Rix, H.-W., {et~al.} 2014, The Astrophysical
  Journal, 791, 17

\bibitem[{Maseda {et~al.}(2016)Maseda, van~der Wel, Rix, Momcheva, Brammer,
  Franx, Lundgren, Skelton, \& Whitaker}]{Maseda2016}
---. 2016, 1

\bibitem[{Momcheva {et~al.}(2016)Momcheva, Brammer, van Dokkum, Skelton,
  Whitaker, Nelson, Fumagalli, Maseda, Leja, Franx, Rix, Bezanson, Cunha,
  Dickey, Schreiber, Illingworth, Kriek, Labb{\'{e}}, Lange, Lundgren, Magee,
  Marchesini, Oesch, Pacifici, Patel, Price, Tal, Wake, van~der Wel, \&
  Wuyts}]{Momcheva2016}
Momcheva, I.~G., Brammer, G.~B., van Dokkum, P.~G., {et~al.} 2016, The
  Astrophysical Journal Supplement Series, 225, 27

\bibitem[{Nakajima {et~al.}(2016)Nakajima, Ellis, Iwata, Inoue, Kusakabe,
  Ouchi, \& Robertson}]{Nakajima2016}
Nakajima, K., Ellis, R.~S., Iwata, I., {et~al.} 2016, The Astrophysical
  Journal, 831, L9

\bibitem[{Nakajima \& Ouchi(2014)}]{Nakajima2014}
Nakajima, K., \& Ouchi, M. 2014, Monthly Notices of the Royal Astronomical
  Society, 442, 900

\bibitem[{Nanayakkara {et~al.}(2016)Nanayakkara, Glazebrook, Kacprzak, Yuan,
  Tran, Spitler, Kewley, Straatman, Cowley, Fisher, Labbe, Tomczak, Allen, \&
  Alcorn}]{Nanayakkara2016}
Nanayakkara, T., Glazebrook, K., Kacprzak, G.~G., {et~al.} 2016, The
  Astrophysical Journal, 828, 21

\bibitem[{Oke \& Gunn(1983)}]{Oke1983}
Oke, J.~B., \& Gunn, J.~E. 1983, The Astrophysical Journal, 266, 713

\bibitem[{Ouchi {et~al.}(2005)Ouchi, Shimasaku, Akiyama, Sekiguchi, Furusawa,
  Okamura, Kashikawa, Iye, Kodama, Saito, Sasaki, Simpson, Takata, Yamada,
  Yamanoi, Yoshida, \& Yoshida}]{Ouchi2005}
Ouchi, M., Shimasaku, K., Akiyama, M., {et~al.} 2005, The Astrophysical
  Journal, 620, L1

\bibitem[{Papovich {et~al.}(2010)Papovich, Momcheva, Willmer, Finkelstein,
  Finkelstein, Tran, Brodwin, Dunlop, Farrah, Khan, Lotz, McCarthy, McLure,
  Rieke, Rudnick, Sivanandam, Pacaud, \& Pierre}]{Papovich2010}
Papovich, C., Momcheva, I., Willmer, C. N.~A., {et~al.} 2010, The Astrophysical
  Journal, 716, 1503

\bibitem[{Quadri \& Williams(2010)}]{Quadri2010}
Quadri, R.~F., \& Williams, R.~J. 2010, The Astrophysical Journal, 725, 794

\bibitem[{Reddy {et~al.}(2012)Reddy, Pettini, Steidel, Shapley, Erb, \&
  Law}]{Reddy2012}
Reddy, N.~a., Pettini, M., Steidel, C.~C., {et~al.} 2012, The Astrophysical
  Journal, 754, 25

\bibitem[{Robertson {et~al.}(2015)Robertson, Ellis, Furlanetto, \&
  Dunlop}]{Robertson2015}
Robertson, B.~E., Ellis, R.~S., Furlanetto, S.~R., \& Dunlop, J.~S. 2015, The
  Astrophysical Journal, 802, L19

\bibitem[{Robertson {et~al.}(2013)Robertson, Furlanetto, Schneider, Charlot,
  Ellis, Stark, McLure, Dunlop, Koekemoer, Schenker, Ouchi, Ono, Curtis-Lake,
  Rogers, Bowler, \& Cirasuolo}]{Robertson2013}
Robertson, B.~E., Furlanetto, S.~R., Schneider, E., {et~al.} 2013, The
  Astrophysical Journal, 768, 71

\bibitem[{Salmon {et~al.}(2015)Salmon, Papovich, Finkelstein, Tilvi, Finlator,
  Behroozi, Dahlen, Dav{\'{e}}, Dekel, Dickinson, Ferguson, Giavalisco, Long,
  Lu, Mobasher, Reddy, Somerville, \& Wechsler}]{Salmon2015}
Salmon, B., Papovich, C., Finkelstein, S.~L., {et~al.} 2015, The Astrophysical
  Journal, 799, 183

\bibitem[{Salzer {et~al.}(2005)Salzer, Lee, Melbourne, Hinz, Alonso‐Herrero,
  \& Jangren}]{Salzer2005}
Salzer, J.~J., Lee, J.~C., Melbourne, J., {et~al.} 2005, The Astrophysical
  Journal, 624, 661

\bibitem[{Sanders {et~al.}(2016)Sanders, Shapley, Kriek, Reddy, Freeman, Coil,
  Siana, Mobasher, Shivaei, Price, \& de~Groot}]{Sanders2016}
Sanders, R.~L., Shapley, A.~E., Kriek, M., {et~al.} 2016, The Astrophysical
  Journal, 825, L23

\bibitem[{Sargent \& Searle(1970)}]{Sargent1970}
Sargent, W. L.~W., \& Searle, L. 1970, The Astrophysical Journal, 162, L155

\bibitem[{Scoville {et~al.}(2007)Scoville, Aussel, Brusa, Capak, Carollo,
  Elvis, Giavalisco, Guzzo, Hasinger, Impey, Kneib, LeFevre, Lilly, Mobasher,
  Renzini, Rich, Sanders, Schinnerer, Schminovich, Shopbell, Taniguchi, \&
  Tyson}]{Scoville2007}
Scoville, N., Aussel, H., Brusa, M., {et~al.} 2007, The Astrophysical Journal
  Supplement Series, 172, 1

\bibitem[{Skelton {et~al.}(2014)Skelton, Whitaker, Momcheva, Brammer, van
  Dokkum, Labb{\'{e}}, Franx, van~der Wel, Bezanson, {Da Cunha}, Fumagalli,
  {F{\"{o}}rster Schreiber}, Kriek, Leja, Lundgren, Magee, Marchesini, Maseda,
  Nelson, Oesch, Pacifici, Patel, Price, Rix, Tal, Wake, \&
  Wuyts}]{Skelton2014}
Skelton, R.~E., Whitaker, K.~E., Momcheva, I.~G., {et~al.} 2014, The
  Astrophysical Journal Supplement Series, 214, 24

\bibitem[{Smit {et~al.}(2014)Smit, Bouwens, Labb{\'{e}}, Zheng, Bradley,
  Donahue, Lemze, Moustakas, Umetsu, Zitrin, Coe, Postman, Gonzalez,
  Bartelmann, Ben{\'{i}}tez, Broadhurst, Ford, Grillo, Infante, Jimenez-Teja,
  Jouvel, Kelson, Lahav, Maoz, Medezinski, Melchior, Meneghetti, Merten,
  Molino, Moustakas, Nonino, Rosati, \& Seitz}]{Smit2014}
Smit, R., Bouwens, R.~J., Labb{\'{e}}, I., {et~al.} 2014, The Astrophysical
  Journal, 784, 58

\bibitem[{Spitler {et~al.}(2012)Spitler, Labb{\'{e}}, Glazebrook, Persson,
  Monson, Papovich, Tran, Poole, Quadri, van Dokkum, Kelson, Kacprzak,
  McCarthy, Murphy, Straatman, \& Tilvi}]{Spitler2012}
Spitler, L.~R., Labb{\'{e}}, I., Glazebrook, K., {et~al.} 2012, The
  Astrophysical Journal, 748, L21

\bibitem[{Stark {et~al.}(2014)Stark, Richard, Siana, Charlot, Freeman, Gutkin,
  Wofford, Robertson, Amanullah, Watson, \& Milvang-Jensen}]{Stark2014}
Stark, D.~P., Richard, J., Siana, B., {et~al.} 2014, Monthly Notices of the
  Royal Astronomical Society, 445, 3200

\bibitem[{Steidel {et~al.}(1998)Steidel, Adelberger, Dickinson, Giavalisco,
  Pettini, \& Kellogg}]{Steidel1998}
Steidel, C.~C., Adelberger, K.~L., Dickinson, M., {et~al.} 1998, The
  Astrophysical Journal, 492, 428

\bibitem[{Straatman {et~al.}(2014)Straatman, Labb{\'{e}}, Spitler, Allen,
  Altieri, Brammer, Dickinson, van Dokkum, Inami, Glazebrook, Kacprzak,
  Kawinwanichakij, Kelson, McCarthy, Mehrtens, Monson, Murphy, Papovich,
  Persson, Quadri, Rees, Tomczak, Tran, \& Tilvi}]{Straatman2014}
Straatman, C. M.~S., Labb{\'{e}}, I., Spitler, L.~R., {et~al.} 2014, The
  Astrophysical Journal, 783, L14

\bibitem[{Straatman {et~al.}(2016)Straatman, Spitler, Quadri, Labb{\'{e}},
  Glazebrook, Persson, Papovich, Tran, Brammer, Cowley, Tomczak, Nanayakkara,
  Alcorn, Allen, Broussard, van Dokkum, Forrest, van Houdt, Kacprzak,
  Kawinwanichakij, Kelson, Lee, McCarthy, Mehrtens, Monson, Murphy, Rees,
  Tilvi, \& Whitaker}]{Straatman2016}
Straatman, C. M.~S., Spitler, L.~R., Quadri, R.~F., {et~al.} 2016, The
  Astrophysical Journal, 830, 51

\bibitem[{Trainor {et~al.}(2016)Trainor, Strom, Steidel, \&
  Rudie}]{Trainor2016}
Trainor, R.~F., Strom, A.~L., Steidel, C.~C., \& Rudie, G.~C. 2016, The
  Astrophysical Journal, 832, 171

\bibitem[{van~der Wel {et~al.}(2011)van~der Wel, Straughn, Rix, Finkelstein,
  Koekemoer, Weiner, Wuyts, Bell, Faber, Trump, Koo, Ferguson, Scarlata, Hathi,
  Dunlop, Newman, Dickinson, Jahnke, Salmon, de~Mello, Kocevski, Lai, Grogin,
  Rodney, Guo, McGrath, Lee, Barro, Huang, Riess, Ashby, \&
  Willner}]{VanderWel2011}
van~der Wel, A., Straughn, a.~N., Rix, H.-W., {et~al.} 2011, The Astrophysical
  Journal, 742, 111

\bibitem[{van~der Wel {et~al.}(2012)van~der Wel, Bell, H{\"{a}}ussler, McGrath,
  Chang, Guo, McIntosh, Rix, Barden, Cheung, Faber, Ferguson, Galametz, Grogin,
  Hartley, Kartaltepe, Kocevski, Koekemoer, Lotz, Mozena, Peth, \&
  Peng}]{VanderWel2012}
van~der Wel, A., Bell, E.~F., H{\"{a}}ussler, B., {et~al.} 2012, The
  Astrophysical Journal Supplement Series, 203, 24

\bibitem[{Vanzella {et~al.}(2009)Vanzella, Giavalisco, Dickinson, Cristiani,
  Nonino, Kuntschner, Popesso, Rosati, Renzini, Stern, Cesarsky, Ferguson, \&
  Fosbury}]{Vanzella2009}
Vanzella, E., Giavalisco, M., Dickinson, M., {et~al.} 2009, The Astrophysical
  Journal, 695, 1163

\end{thebibliography}

\end{document}